\documentclass[runningheads]{llncs}
\usepackage{mycommands}

\begin{document}
\title{A Constraint Opinion Model
\thanks{This work has been partially supported by the SGR project PROMUEVA (BPIN 2021000100160) under the supervision of Minciencias (Ministerio de Ciencia
Tecnolog\'ia e Innovaci\'on, Colombia), by the MUR project 
STENDHAL (PRIN 20228KXFN2), by the CNRS project TOBIAS under the MITI interdisciplinary programs, and by the NATO project SymSafe
(Science for Peace and Security Programme G6133)
}}

\author{Fabio Gadducci\inst{1}
\and
Carlos Olarte\inst{2}
\and
Frank Valencia\inst{3,4}
}
\authorrunning{F. Gadducci et al.}
\institute{
Department of Computer Science, University of Pisa, Italy
\email{fabio.gadducci@unipi.it} \and
 LIPN, CNRS UMR 7030, Universit\'e Sorbonne Paris Nord,  France 
\email{olarte@lipn.univ-paris13.fr} \and
CNRS-LIX, École Polytechnique de Paris, France \and
Department of Electronics and Computer Science, Pontificia Universidad Javeriana Cali, Colombia
\email{frank.valencia@lix.polytechnique.fr}
}
\maketitle              \begin{abstract}
This paper introduces a generalised opinion model that extends the standard
DeGroot model by representing agents' opinions and influences as soft
constraints rather than single real values. This allows for modelling scenarios
beyond the scope of the DeGroot model, such as agents sharing partial
information and preferences, engaging in discussions on multiple topics
simultaneously, and representing opinions with different degrees of
uncertainty. By considering soft constraints as influences, the proposed model
captures also situations where agents impose conditions on how others' opinions
are integrated during belief revision. 
Finally, the flexibility offered by soft constraints allows us to introduce a novel
polarisation measure that takes advantage of this generalised framework.

\end{abstract}
\section{Introduction}\label{sec:intro}
Social networks play a significant role in \emph{opinion formation}, consensus building, and polarisation among their users.  
The dynamics of opinion formation in such networks typically involves
individuals sharing their views with their contacts, encountering different
perspectives, and adjusting their beliefs in response. Models of opinion
formation
\cite{Degroot1974,demarzo2003persuasion,AlvimAKQV23,Aranda2024} capture these
dynamics to simulate and reason about opinion evolution.

The DeGroot model \cite{Degroot1974} is one of the most representative
formalisms for opinions' formation and consensus' building in social networks. In
this model, a social network is represented as a directed \emph{influence
graph}, whose edges denote the weight, expressed as a \emph{real number}, that an
\emph{agent} (i.e., an individual) carries on another. Each agent has
an opinion, also expressed as \emph{real number}, indicating the level of
agreement with an underlying proposition. Agents repeatedly update their
opinions by taking the weighted average of the opinions of those who influence
them (i.e., their \emph{neighbors} or \emph{contacts}). 

The DeGroot model
is widely recognised for its elegant characterisation of opinion consensus
based on the topology of the influence graph, and it remains a central focus of
research for developing frameworks to understand opinion formation dynamics in
social networks
(e.g.~\cite{DBLP:conf/forte/AlvimAKQV21,fc,demarzo2003persuasion,chatterjee1977towards,paz:hal-04918975,AlvimAKQV23,Aranda2024}).
Nevertheless, when modelling real scenarios from social networks, we often only have  \emph{partial information} about agents' opinions and the influence they have on one another.  In practice, opinion information may be incomplete or imprecise due to privacy constraints, self-censorship, or the dynamic nature of beliefs. Similarly, influence relationships are often difficult to quantify, as they depend on factors such as trust, authority, and exposure to diverse viewpoints, which are not always explicitly observable. This \emph{uncertainty} hinders the application of classical models like DeGroot, which assume for all agents fully known opinions and influences, typically represented as real values.

In this paper we introduce \emph{Constraint Opinion Models}, a framework where
both opinions and influences are represented as \emph{(soft) constraints} rather
than exact real values. This allows us to reason about
meaningful situations where only partial information or preferences are
available. We thus  generalise the DeGroot model, at the level of opinions and influences, 
while keeping much of
its mathematical simplicity.

We show that using soft constraints to represent opinions offers several advantages.
First, they seamlessly represent opinions on different topics or propositions (i.e., \emph{multi-dimensional} opinions),
enabling the analysis of network behavior as agents discuss various subjects.
Second, they allow for the representation of \emph{uncertainty} and \emph{partial
information}. This is particularly important to model situations where an agent's
exact opinion is unknown. Third, they support some forms of  \emph{epistemic
modeling}, capturing beliefs where agents hold opinions about other agents.
Additionally, the framework can express complex opinions that a single value
cannot adequately represent, such as ``extreme'' viewpoints (e.g., agents that prefer any extreme option than a moderate one).

Regarding the representation of influences, soft constraints provide also an extra
flexibility. We will show that they enable the definition of ``filters'' that impose boundaries
on how agents adjust their opinions while preserving their core beliefs.
Moreover, soft constraints allow for representing conditional influences,
where the weight of the influence depends on the incoming information or the
subject being discussed.

Finally, a key challenge in social system analysis is measuring the difference between
two opinions as this is the basis of any \emph{polarisation} measure \cite{ER:1994}. In the DeGroot model, where opinions are real values, this is
straightforward. To extend this capability to constraint-based opinion models,
we introduce a notion of distance between constraints. This allows for
quantifying opinion divergence and assessing how polarised a system of agents
sharing constraints becomes.

\paragraph{Organization.} After recalling the basic notions of opinion models
and semiring-based constraints in \Cref{sec:prelim}, we define constraint
opinion models in \Cref{sec:co-models}. 
\Cref{sec:app} is devoted to illustrate with several
examples the possibilities offered by our framework in modelling scenarios beyond
the classical DeGroot opinion model. Our novel notion of 
distance between soft constraints is given in \Cref{sec:polarization}. 
\Cref{sec:concluding} concludes the paper. The
experiments shown in \Cref{sec:app} can be reproduced with the aid of a tool 
available at \url{https://github.com/promueva/constraint-opinion-model}. 
 
\section{Preliminaries}\label{sec:prelim}
This section recalls some results on semirings, which are the algebraic
structures adopted here for modeling (soft) constraints (\Cref{sec:lem}). We
also recall the notion of opinion models and belief revision in the standard
DeGroot model (\Cref{sec:odm}).

\subsection{Monoids, Semirings and Soft Constraints}\label{sec:lem}

We shall use the values of a monoid, which can be combined, to represent \emph{preferences}.

\begin{definition}[Monoids, groups]\label{defn:clm}
	A (commutative) monoid is a triple
	$\langle A, \oplus, \0 \rangle$ such that $\oplus: A \times A \rightarrow A$ is
	a commutative and associative function and $\0 \in A$ its \emph{identity} element,
	i.e. $\forall a \in A. a \oplus \0 = a$. 
A group is a four-tuple
	$\langle A, \oplus, \ominus, \0 \rangle$ such that 
    $\langle A, \oplus, \0 \rangle$ is a monoid and 
    $\ominus: A \times A \rightarrow A$
	a function satisfying $\forall a \in A. a \ominus a = \0$. 
\end{definition}

As usual, we use the infix notation: $a \monop b$ stands for $\monop(a,b)$.

\begin{definition}[Semirings, rings]\label{defn:cls}
	A semiring $\bbS$ is a five-tuple
	$\langle A, \oplus, \0, \monop, \1 \rangle$ such that 
	$\langle A, \oplus, \0 \rangle$ and 
	$\langle A, \monop, \1 \rangle$ are monoids
	satisfying  an annihilation law, i.e. 
    $\forall a \in A. a \otimes \0 = \0$, 
    and a distributive law, i.e.
	$\forall a, b, c \in A. a \otimes (b \oplus c) = (a \otimes b) \oplus (a \otimes c)$.
    A ring is a six-tuple
    $\langle A, \oplus, \ominus, \0, \monop, \1 \rangle$
such that $\langle A, \oplus, \0, \monop, \1 \rangle$ is a semiring and $\langle A, \oplus, \ominus, \0 \rangle$ a group.
\end{definition}

\begin{remark}
    
In the soft constraint tradition~\cite{jacm,jlamp17}, 
it suffices to consider semirings, and often $\oplus$ is actually idempotent, hence resulting in a tropical semiring.
However, for some of our examples we will resort to rings and, sometimes, to fields, i.e. where there is an additional binary operation $\odiv$ that is defined whenever the second operand is different from $\0$ and such that $a \odiv a = \1$.
\end{remark}

\paragraph{Soft Constraints.} We now introduce \emph{soft}
constraints that generalise classical constraints: in the latter,
a variable  assignment  $\eta$ satisfies or not a constraint, while in
the former, $\eta$  is assigned to a semiring value, interpreted as the level
of preference, importance, fuzziness, cost, uncertainty, etc.  of such assignment. Our
definition is a straightforward generalisation of the one adopted for
optimisation problems~\cite{jacm,jlamp17}, where $\oplus$ is idempotent. We fix
a semiring $\mathbb{S} = \langle S, \oplus, \0, \monop, \monid \rangle$.

\begin{definition}[Soft constraints]\label{def:softconstraints}
	Let $V$ be a set of variables and $D$ a finite domain of interpretation.
	A \emph{(soft) constraint} over $\mathbb{S}$ is a function $c: (V \rightarrow D) \rightarrow
	S$ associating a value in $S$ for each variable assignment (or valuation) 
    $\eta: V\rightarrow D$ of the variables. ${C}$ is the set
    of all the possible constraints that can be built from $S$, $D$, and $V$. 
\end{definition}

Let $\eta:V\rightarrow D$ be a valuation 
and $c:(V \rightarrow D) \rightarrow S$ a constraint. 
We use $c\eta$ to denote the semiring value obtained when $c$ is applied to $\eta$.
With $\eta[x:=d]$ we denote the valuation $\eta'$ 
where $\eta'(y) = \eta(y)$ for all $y\in V\setminus\{x\}$
and $\eta'(x) = d$. Given a set $X\subseteq V$ of variables, 
we use $\eta\downarrow_X=\eta'\downarrow_X$
to denote the fact that 
 $\eta(x) = \eta'(x)$ for all $x\in X$. 

We use  $c^{-1}: S \to 2^{V \to D}$ to denote the ``inverse'' of a constraint
$c$, i.e., $c^{-1}(s)$ is the set of assignments $\Xi=\{\eta~\mid~c\eta = s\}$.
We say that the set $\Xi$ is the set of \emph{solutions} of $c$ with respect to
$s$ (in the sense that they map $c$ into a designated value $s$).

A constraint $c$ often depends only on a subset of the variables in $V$.
Formally, a constraint $c$ depends on the set of variables $X \subseteq V$ 
if for all valuations  $\eta,\eta'$ we have   $c\eta = c\eta'$
whenever $\eta\downarrow_X=\eta'\downarrow_X$.
The smallest such set 
is called the
\emph{support} of $c$ and denoted as $sv(c)$:
It identifies the \emph{relevant} variables of the constraint
$c$. 
In fact, if $sv(c) = \{x_1,\cdots, x_n\}$, we often write $c[x_1\mapsto
v_1,\cdots,x_n\mapsto v_n]=s$ to denote the fact that $c\eta = s$ for any
$\eta$ that maps each $x_i$ into $v_i$, since the assignment to other
variables is irrelevant. 

The projection of a constraint $c$ on a set of variables $X$, denoted as $\projC{c}{X}$, is
the constraint $c'$ such that  $c'\eta = \bigoplus_{\{\eta'\mid \eta'\downarrow_X = \eta\downarrow_X \}}(c\eta')$. This means that the 
variables \emph{not} in $X$ are 
 ``removed'' from the support of $c$ in $c'$.

We call $c$ a
\emph{constant} constraint if there exists $s$ such that  $c\eta = s$ for all valuation $\eta$ 
 (and hence, $sv(c)=\emptyset$). With a slight abuse of notation, we shall identify a constant
constraint $c$, where $c\eta=s$, with the semiring value $s$.

The set of constraints ${C}$ forms a semiring $\mathbb{C}$, whose structure is lifted
from ${\mathbb S}$. More precisely, $(c_1 \star c_2)\eta = c_1\eta \star
c_2\eta$ for all $\eta: V \rightarrow D$ and  $\star \in \{\oplus,
\otimes\}$. Combining constraints by the monoidal operators means building a
new constraint whose support involves, at most, the variables of the original
ones, since it is easily proved that 
$sv(c_1 \star c_2) \subseteq sv(c_1) \cup
sv(c_2)$. The resulting constraint is associated with each tuple of domain values
for such variables, which is the element that is obtained by adding/multiplying
those associated with the original constraints to the appropriate sub-tuples.

\begin{example}\label{ex:sem}

	Consider the Boolean semiring $\bbB=\langle \{\TT,\FF\},\vee,\FF ,\wedge,\TT\
\rangle$, a set of variables $V$
and the
integer domain $D = [0,100]$. 
The set of constraints $C$ whose support is contained in the subset $\{x,y\} \subseteq V$ includes the
constant constraints $\TT$ and $\FF$, constraints such as $c_1 = \{x \leq 42\}$ (i.e.,
$c_1[x\mapsto v]=\TT$ iff  $0\leq v \leq 42$) and $c_2=\{y \leq 25\}$, as well as
their compositions $c_1 \vee c_2$ and $c_1 \wedge c_2$. As expected,
$(c_1\wedge c_2)[x\mapsto v_x, y\mapsto v_y]=\TT$ iff $v_x\leq 42$ and $v_y
\leq 25$. Moreover, $\projC{(c_1\wedge c_2)}{\{x\}} = c_1$. 

\end{example}

\subsection{Opinion Dynamic Models}\label{sec:odm}

This section recalls the notion of belief update \`{a} la DeGroot \cite{Degroot1974}. The
definitions below are taken from \cite{DBLP:conf/forte/AlvimSKV24,paz:hal-04918975}.

\begin{definition}[Influence graph]\label{def:inf-graph}
An ($n$-agent) \emph{influence graph} is a directed graph 
$G = \langle A, E, I \rangle$ such that $A$
is the set of vertices, 
$E \subseteq A \times A$ the set of edges, and 
$I : E \to [0, 1]$ the weight function.
\end{definition}

In the following, the set $A$ of vertices
is always given by an interval
$\{1,  \ldots , n\}$ of integers, and
$I$ is extended to $I : A \times A \to [0, 1]$ assuming that 
$I(i, j) = 0$ if $(i, j) \not \in E$.

The vertices in $A$ represent $n$ agents of a community or network. The edges
$E$ represent the (direct) influence relationship between these agents, i.e. $(i,
j) \in E$ means that agent $i$ influences agent $j$. The value $I(i, j)$
denotes the strength of the influence, where a higher
value means a stronger influence.

As expected, the graph $G$ can be represented as a square matrix $M_G$ of $n=|A|$
rows where $M_{j,i}=I(i,j)$ (i.e., $M_{j,i}$ is the degree of influence
of agent $i$ on agent $j$).
Hence, we shall identify the graph $G$ with its corresponding matrix $M_G$ and
omit the subindex $G$ in $M_G$ when the influence graph can be deduced from the context. 
We denote by $A_i$ the set $\{j \mid (j, i) \in E\}$ (i.e., $\{j \mid M_{i,j}\neq 0\}$) of agents with a direct
influence over agent $i$.

At each time unit $t$, all the agents update their opinions. We use $B^t:A\to[0,1]$ (that
can be seen as a vector of $|A|$ elements)  to
denote the state of opinion at time $t$, and $B^t_i$ to denote the opinion of
agent $i$ at time-unit $t$. A DeGroot-like opinion model describes the
evolution of agents' opinions about some underlying statement or proposition.

\begin{definition}[Opinion model]\label{def:o-model} An \emph{opinion model} is a triple $\langle
    G, B^0, \mu_G\rangle$ where $G = \langle A, E, I \rangle$ is an n-agent influence graph, $B^0: A \to [0, 1]$ is the
    initial state of opinion, and $\mu_G : [0, 1]^n \to [0, 1]^n$ is a
    state-transition function, called update function. For every $t$, the state
    of opinion at time $t + 1$ is given by $B^{t+1} = \mu_G(B^t)$. 
\end{definition}

The update function $\mu_G$ is dependent on $G$ but can be tuned to model different cognitive
biases~\cite{DBLP:conf/forte/AlvimSKV24,paz:hal-04918975}, including e.g.
\emph{confirmation bias} (where agents are more receptive to opinions that
align closely with their own), the \emph{backfire effect} (where agents
strengthen their position of disagreement in the presence of opposing views),
and \emph{authority bias} (where individuals tend to follow authoritative or
influential figures, often to an extreme). In this paper we  focus on the following 
biased update function from~\cite{paz:hal-04918975}
\begin{equation}\label{eq:update2}
B^{t+1}_i  = B^t_i + R_i \sum_{j \in A_i}\beta^t_{i,j} M_{i,j} (B^t_j - B^t_i)
\end{equation}
where $R_i = \frac{1}{\sum_j M_{i,j}}$ 
if $\sum_j M_{i,j} \neq 0$ and $0$ otherwise, 
and $\beta_{i,j}^t=\beta_{i,j}(B^t_i,B^t_j)$ is a value in $[0,1]$ possibly depending on
$B^t_i$ and $B^t_j$.

A broad class of update functions, generalizing 
the 
DeGroot model, can be obtained from \Cref{eq:update2}. 
Intuitively, updates for an agent $i$ can \emph{weight}
disagreements $(B_j - B_i)$ with each one of its neighbors $j$ using functions
$\beta_{i,j}$ from the opinion  states of $i$ and $j$ to $[0,1]$. These functions are referred to as
\emph{(generalised) bias factors}.
Notice that the same opinion difference can then be weighted differently by bias
factors, depending of current opinions of agents $i$ and $j$. Thus,
intuitively $\beta_{i,j}$ may also be seen as \emph{dynamically changing} the
constant influence of $j$ over $i$, i.e.,  $M_{i,j}$, depending on their opinions.

\begin{remark}
If  $\beta^t_{i,j}=1$, \Cref{eq:update2} corresponds to the update function in
the  classical DeGroot model. In~\cite{DBLP:conf/forte/AlvimSKV24} it is
studied the case for $\beta^t_{i,j} = 1 - \mid B^t_j - B^t_i \mid$, which is
introduced as the confirmation bias factor of $i$ with respect to $j$ at time
$t$. It is not difficult to prove that in both cases, we always have that
$B^{t+1} \in [0,1]^n$.
\end{remark}

It should be noted that the formula in \Cref{eq:update2} can be obtained by
manipulating suitable matrices as shown in the following remark. 

\begin{remark}
\label{calcul}
Consider the vector $\1$ such that 
$\1_i = 1$, the vector 
$R$ such that 
$R_i = \frac{1}{(M\1)_i}$
if $(M\1)_i \neq 0$
and $0$ otherwise,
the matrices $U^t$ obtained as the Hadamard product of $\beta^t$ and $M$, i.e. $U^t_{i,j} = 
\beta^t_{i,j} M_{i,j}$, and the diagonal matrices 
\begin{gather*}
	\small
N = diag(R) = \begin{bmatrix}
		R_1  & 0 & \ldots & 0  \\
		0 & R_2 & \ldots & 0 \\
        \ldots & \ldots & \ldots & \ldots\\
        0 & 0 & \ldots & R_n 
	\end{bmatrix}
\qquad
	\small
V^t = diag(U^t \1) =
   \begin{bmatrix}
		\sum_j U^t_{1,j}  & 0 & \ldots & 0  \\
		0 & \sum_j U^t_{2,j} & \ldots & 0 \\
        \ldots & \ldots & \ldots & \ldots\\
        0 & 0 & \ldots & \sum_j U^t_{n,j}
	\end{bmatrix}
\end{gather*}
It is now easy to see that 
\[
B^{t+1}_i  = B^t_i + N_{i,i} \sum_{j} U^t_{i,j} (B^t_j
- B^t_i) = B^t_i + N_{i,i} ((\sum_{j} U^t_{i,j} B^t_j)
- (\sum_{j} U^t_{i,j} B^t_i)) = 
\]
\[
= B^t_i + N_{i,i}
((U^t B^t)_i - (V^t_{i,i} B^t_i))
\]
so that, for $I$ the diagonal identity matrix, we have
\begin{gather}
\small
B^{t+1} = 
B^t + N (U^t B^t - V^t B^t) =
(I + N(U^t - V^t)) B^t
\end{gather}
where 
$(U^t - V^t) \1 = 0$, and hence, 
$(I + N(U^t - V^t)) \1 = \1$. In other terms, $I + N(U^t - V^t)$
is a \emph{row stochastic} matrix, where the sum of the elements in each row is $1$.

In the case of $\beta_{i,j} = 1$, we have that $U^t = M$ and $V^t = diag(M\1)$, so that $NV = I$
and
$B^{t+1} = NMB^t$.
If $M$ is also assumed to be a
row stochastic matrix, i.e. it satisfies $M\1 = \1$, then $N = I$
and thus $B^{t+1}=MB^t$,
as expected in the  DeGroot model. \qed 
\end{remark}

Given a matrix $M$ representing the influence graph $G$ of an opinion model, we shall use $M^*$ to denote the limit of $M^t$
for 
$t\to\infty$.  If $M$ is row stochastic and, furthermore, \emph{strongly connected} (i.e. the influence
graph it represents is strongly connected) and \emph{aperiodic} (i.e. the greatest
common divisor of the lengths of its cycles is one)~\cite{Golup2017}, the limit
$M^*$ exists and the rows of $M^*$ are all the same. 
In this case, for all pair
of agents $i$ and $j$ and initial beliefs $B$, $(M^*B)_i = (M^*B)_j$. Thus, assuming $\beta_{i,j} = 1$, so that $B^{t+1}=MB^t$ as in the DeGroot model, we obtain that, in the limit, the opinion of the agents converges to the same value, thus reaching a
\emph{consensus} independently from their initial beliefs. A simple and sufficient
condition for $M$ to be aperiodic is by checking that there exists an agent $i$
such that $M_{i,i}>0$. An agent where $M_{i,i}=0$ can be seen as a \emph{puppet}
that just incorporates the opinions of others.

\begin{theorem}[Consensus in the DeGroot model~\cite{Degroot1974}]\label{prop:conc} 
Let $M$ be a row stochastic
matrix of values in $[0,1]$. If  $M$ is strongly connected and
aperiodic, then 
$\lim_{t\to\infty}M^t$
exists
and all its rows 
are equal.
\end{theorem}

A consensus result for more general bias factors $\beta_{i,j}$ under some suitable conditions (including continuity) can be found in \cite{paz:hal-04918975}.

\begin{example}[DeGroot]\label{ex:dgroot}
    Agents $1$ and $2$ discuss about a proposition $p$, and the initial state
    of opinions is $[0.3\ \  0.6]$, where Agent $1$ tends to believe that  $p$
    is not the case while Agent $2$ is more positive about $p$. Consider the
    influence graph 
    $ \small M = 
    \begin{bmatrix} 
        1 & 0  \\ 0.8 & 0.2
    \end{bmatrix} 
    $, which is a row stochastic, strongly connected, and
    aperiodic matrix. According to $M$, Agent $1$ accepts no influence, while Agent
    $2$ does. Recall that in the DeGroot model, $B^{t+1}=MB^t$. Hence,  $B^{1}
    = MB^0 = [0.3\ \ 0.36]$. Since $M_{1,2}=0$, Agent $1$ does not change her
    opinion. Moreover, due to the influence of Agent $1$, the opinion of Agent
    $2$ progressively  converges  to $0.3$, since $\lim_{t\to\infty}M^t=
        \begin{bmatrix} 1 & 0  \\ 1 & 0 \end{bmatrix} $. 
\end{example}

\section{Constraint Opinion Models}\label{sec:co-models}
Opinion models, as shown in the previous section, represent opinions and
influences as a real number in the interval $[0,1]$. This section generalises
this idea and proposes a \emph{constraint opinion model}, where the opinions and the influences of the agents are represented as (soft) constraints in a given
semiring. 

Recall from \Cref{sec:lem} that 
 given a set of variables $V$, a finite domain  $D$,  and a 
semiring $\mathbb{S}$, 
the set of constraints built from 
$\mathbb{S}$, $D$ and $V$ is denoted $C$. In fact, we have a semiring $\mathbb{C}$ of constraints, with carrier $C$,
whose structure is lifted from $\mathbb{S}$.

\begin{definition}[Constraint influence graph]
Let $\mathbb{S}$ be a semiring and $C$ the constraints built from 
$\mathbb{S}$, $V$ba set of variables, and $D$ a finite domain. 
Let $A=\{1,\cdots,n\}$ be a set of agents. 
An ($n$-agents) \emph{constraint influence graph} is a square matrix $M$ of
dimension $n$ with values in $C$, i.e. $M : A \times A \to C$.
\end{definition}

An element $c_{i,j}$ of $M$ (row $i$, column $j$) represents how agent $j$
influences agent $i$. As shown below, this
influence can be a constant or an arbitrary constraint, reflecting scenarios
where agents impose specific restrictions on the way they are influenced.

\begin{definition}[Constraint Opinion Model]\label{def:co-model}
A \emph{constraint opinion model} is a triple $\langle M, B^0, \mu^t\rangle$
where $M$ is an $n$-agent constraint influence graph, 
$B^0: A \to C$ the initial state of opinion, 
and $\mu^t : C^n \to C^n$ the update function at time $t$, so that 
the state of opinion at time $t + 1$ is given by $B^{t+1} = \mu^t(B^t)$.
\end{definition}

As shown in the forthcoming sections, representing the opinion of an agent at
time $t$ as a constraint $B^t_i$ widens the spectrum of situations that can be
modeled in systems of agents updating their beliefs. 
We note that an opinion model (\Cref{def:o-model}) is an instance of a constraint
opinion model where the set of variables $V$ is $\{p\}$, the domain of the
variable $p$ is $D=\{0,1\}$, and the semiring is given by the positive real numbers
$\langle \mathbb{R}^+,+,0,\times, 1 \rangle$. Moreover, any opinion $c$ is required to
satisfy $\sum_\eta c\eta = 1$ (e.g., if $c[p\mapsto 0]=0.3$, then $c[p \mapsto 1]=0.7$), all
the elements in the matrix (influence graph) $M$  are constants, and the update
function has type  $\mu^t : [0,1]^n \to [0,1]^n$. 

In the following examples, it
is assumed that  $\mu^t$ is given by the following matrix multiplication equation
\begin{equation}\label{general-degroot:eq}
    B^{t+1} = \mu^t(B^t) = M B^t = M^t B^0. 
\end{equation}

\begin{example}[Opinion Models in $\bbB$] \label{ex1}
    Consider again the semiring $\bbB$ 
    and the constraints $c_1 = \{x \leq 42\}$ 
    and $c_2=\{y \leq 25\}$ in \Cref{ex:sem}. Let $d_1 = c_1 \wedge c_2 = \{x
    \leq 42, y \leq 25\}$, $d_2 = c_3 \vee c_4$ where $c_3 = \{x \geq 15\}$ and
    $c_4 = \{y \geq 66\}$, and let $B^0=[d_1~ d_2]$ be the initial set of opinions.
    Consider also the following constraint influence graphs 

\begin{gather*}
	\small
	M_1 = \begin{bmatrix}
		\TT &\ & \FF  \\
		\TT &\ & \TT
	\end{bmatrix}\qquad
	M_2 = \begin{bmatrix}
		\TT &\ & \FF  \\
		\FF &\ & \TT
	\end{bmatrix}\qquad
	M_3 = \begin{bmatrix}
		\FF &\ & \TT  \\
		\TT &\ & \FF
	\end{bmatrix}
	\qquad
	M_4 = \begin{bmatrix}
		\TT & \ & {y \leq 20}  \\
		{x \geq 10} &  \ &\TT
	\end{bmatrix}
\end{gather*}

The matrix $M_1$ is idempotent (i.e. $M_1 M_1 = M_1$) and represents the situation where  Agent $1$
is not influenced at all by Agent  $2$ while, instead,
Agent $2$ accepts all the
information from Agent $1$. In this scenario, we have  $M_1 B^0 = B^1 = [d_1~~~~ d_1 \vee
d_2]$, and $M_1 B^1 = B^1$. This means that, after one interaction, the system
stabilizes in an opinion where Agent $1$ does not change its initial opinion
$d_1$, and Agent $2$ considers also possible the values for $x$ and $y$ according
to the opinion $d_1$ of Agent $1$. 

 $M_2$ is the identity matrix $I$,
 and it represents a situation where neither agent is
influenced by the other. Hence,   $M_2 B = B$ for any choice of $B$. Instead, $M_3$ is involutory
(i.e. $M_3 M_3 = I)$
and thus the system never stabilises,
alternating between 
$[d_1~~~ d_2]$
and
$[d_2~~~ d_1]$.

The constraint influence graph $M_4$ is more interesting, since some of the
influences are not constants. Consider the element $(M_4)_{2,1} = \{x \geq
10\}$: Agent $2$ is influenced by Agent $1$ only to the point that
it accepts that $x$ might also be bigger than $10$  (but not smaller than
that). We thus have $M_4 B^0 =
[d_1'\,\,\,\, d_2']$ 
with 
\[
\begin{array}{lll}
d_1' &=& (c_1 \wedge c_2) \vee (\{y \leq 20\} \wedge
(c_3 \vee c_4)) = (c_1 \wedge c_2) \vee \{x \geq 15, y \leq 20\} \\
d_2' &=& (\{x \geq 10\} \wedge c_1 \wedge c_2) \vee (c_3 \vee c_4) = \{ 10 \leq x \leq 42,
y \leq 25\} \vee (c_3 \vee c_4)\\
\end{array} 
\]
The use of constraints in $M_4$ allows us to represent a \emph{core
belief}~\cite{Makinson1997-MAKSR,Hansson-2001}  for Agent $2$, as it
``filters'' (part of) the opinion of Agent $1$ when it is not consistent with
the limits she imposes to  update her opinions. In this case, after
interaction, Agent $2$ still cannot believe that $x=9$, a scenario that Agent
$1$ considers plausible.  This is possible due to the flexibility of using
constraints on the influence graph (and not only on opinions): Agents can
impose different conditions on how they are influenced by other agents.

\end{example}

\begin{example}[Opinion Models in $\bbR^+$]\label{ex:inR}
    Consider the 2-agent influence graph $M$ in \Cref{ex:dgroot}
    and the initial state of opinions $[c_1\ \ c_2]$ where 
 $c_1=\{0 \mapsto 0.3, 1 \mapsto
0.7\}$ and $c_2 = \{0 \mapsto 0.6, 1 \mapsto 0.4\}$, respectively. Remember that according to
$M$, Agent $1$ accepts no influence, while Agent $2$ does. Assume that the update
function  $\mu^t$ is as in \Cref{general-degroot:eq}. We have $MB^0 = [c_1'\ \ c_2']$ with
$c_1' = c_1$ and $c_2' = \{0 \mapsto 0.36, 1\mapsto 0.64\}$. 
Agent $1$ does not change her opinion and,  due to the influence of Agent $1$, the opinion of Agent $2$
progressively  converges  to $c_1$. 
\end{example}

\subsection{On the Update Function}\label{update}

Consider again the calculations in Remark~\ref{calcul}.
Note that they can be mimicked in any field, so that given the Hadamard product 
$U^t = \beta^t \otimes M$ we have 
\begin{gather}
\small
B^{t+1} = 
(I \oplus N(U^t \ominus V^t)) B^t
\end{gather}
Note that in this general case we also have 
$(U^t \ominus V^t) \1 = 0$, 
so that
$(I \oplus N(U^t \ominus V^t)) \1 = \1$, and hence $I \oplus N(U^t \ominus V^t)$ is a row stochastic matrix.
If, moreover, $M$ is also row stochastic, we have
$B^{t+1} = 
(I \oplus U^t \ominus V^t) B^t$,
which is a valid equation for any ring, since the definition of the matrices does not involve the division operator, if $\beta^t$ does not.

\medskip
Let us now consider the two examples above: They adopt the standard DeGroot model, i.e. such that
$B^{t+1} = M B^t$,
which is a valid equation for any semiring.
All the elements of the first three matrices in~\Cref{ex1} are constant constraints.
Also, the four matrices
are strongly connected and row stochastic, since in each row the sum of the constraints gives the constant constraint
always returning $\TT$.
In particular, for $M_4$
we have that 
$\TT \vee {y \leq 20} =
{x \geq 10} \vee \TT = \TT$.
However, $M_3$ is not aperiodic,
since the only cycle has length $2$.
The matrix in~\Cref{ex:inR} contains only constant constraints 
and it is row stochastic, strongly connected, and aperiodic, since each agent has a cycle.

\medskip

The following definition makes precise the idea of a set of agents reaching
a \emph{consensus} (as the two agents in \Cref{ex:inR})
and of a set of agents agreeing in the limit about a particular valuation
for the variables (see \Cref{ex:topics} in the following section). 

\begin{definition}[Consensus]
Let $\langle M, B^0, \mu^t\rangle$ be a
constraint opinion model where $M:A\times A \to C$ is a  constraint 
influence graph.
We say that the set of agents $A$ converges to 
an opinion $c\in C$ whenever for each $i\in A$,
 $\lim_{t\to\infty}B^t_i =c$. 
We say that the set of agents $A$ converges to a consensus whenever each agent in $A$ converges
to the same opinion.
Given a valuation $\eta$, we say
that the agents converge to a consensus about $\eta$
whenever there exists $s\in S$ such that for each $i\in A$, $\lim_{t\to\infty}B^t_i\eta =s$.
\end{definition}

We now move a step further and we consider the set $C_p$ of what we call 
\emph{probability} constraints,
that is, $c\in C_p$ if
$\oplus_\eta c\eta = 1$.

\begin{proposition}
Let $M$ be a row stochastic 
matrix of constant constraints
and $B$ be a vector whose 
elements belong to $C_p$.
Then the elements of the vector $M B$ belong to $C_p$.
\end{proposition}

\section{Sharing and Discussing about Partial Information}\label{sec:app}
 This section illustrates how constraint opinion models provide an extra
expressiveness for representing scenarios involving agents sharing
opinions. We explore situations that cannot be modeled  using the classical 
DeGroot framework, such as agents discussing preferences and exchanging partial
information (\Cref{app:pref}), modeling conditional preferences
(\Cref{app:cond}), and capturing agents' beliefs about one another
(\Cref{app:bel}). In all the examples reported here, the update function
is given as in \Cref{general-degroot:eq}.

\subsection{Preferences and Partial Information}\label{app:pref}
    We consider agents engaging in discussions about a given proposition $p$. As
noted in \Cref{sec:co-models}, when the constraint opinion model is
instantiated to represent the DeGroot model, the domain of $p$ is
restricted to $\{0,1\}$. In the section,  instead, we consider a different
domain for $p$ allowing us to represent situations where agents give different
\emph{preferences} to $p$, or when they only have \emph{partial information}
about $p$. This will be useful to model opinions that cannot be 
represented in the DeGroot model (as a single real number in the interval $[0,1]$). 

Consider a variable $p$ with finite domain $D = \{1,\cdots,5\}$, where $1$
means ``very bad'' and $5$ means ``very good'' as in a Likert scale, or
alternatively, ``extreme-left'' and ``extreme-right'', ``strongly disagree''
and ``strongly agree'', etc.  Let $a,b\in D$ such that $a\leq b$ and define the
constraint $\CInt{a,b}$ as $\CInt{a,b}(v)= \frac{1}{1+(b-a)}$ if $v\in [a,b]$
and $\CInt{a,b}(v)=0$ otherwise. The constraint $\CInt{a,b}$ represents an
opinion where the agent assigns equal preferences (different from 0) when
the value of $p$ is in the interval  $[a,b]$ and $0$ otherwise.
Note that for all $a,b$ such that $1\leq a\leq b\leq 5$, $\CInt{a,b} \in C_p$
(i.e.
$\oplus_\eta \CInt{a,b}\eta = 1$). 

Consider the following inference graphs and initial opinions

\[
	\begin{array}{lllllllllllll}
	M_1 = & \left[
			\begin{array}{lll}
				\CCte{0.9} && \CCte{0.1}\\
				\CCte{0.8} && \CCte{0.2}\\
		
			\end{array}
		  \right]
	& \quad & 
	M_2 = & \left[
			\begin{array}{lll}
				\CCte{0.1} && \CCte{0.9}\\
				\CCte{0.2} && \CCte{0.8}\\
		
			\end{array}
		  \right]
	& \quad & 
	M_3 = & \left[
			\begin{array}{lll}
				\CCte{0.5} && \CCte{0.5}\\
				\CCte{0.5} && \CCte{0.5}\\
		
			\end{array}
		  \right]
	& \quad & 
B_0 = & [ \CInt{1,2} \quad \CInt{1,5} ] 
	\end{array}
\]

Regarding the initial opinions, Agent $1$ is somewhat negative about $p$ but
\emph{uncertain} whether to assess $p$ as ``very bad'' or merely ``bad.'' In
contrast, Agent $2$ has no specific opinion about $p$ and thus assigns an equal
preference $0.2$ to all possible values of $p$. It is worth noticing that the
partial information conveyed by these  opinions cannot be accurately expressed in
the DeGroot model as a single value within the interval $[0,1]$.

When the influence graph  $M_1$ is considered, the system converges to a consensus
where the two agents assign a preference of approximately $0.467$ when $p\in \{1,2\}$ and
$0.022$ otherwise. In this scenario, Agent $1$ has a stronger influence on Agent $2$ and then
 the agents tend to agree that $p$ is ``very bad'' or ``bad''. In the
case of $M_2$, the system converges to a preference of approximately $0.255$ when $p\in \{1,2\}$ and
$0.163$ otherwise: since Agent $2$ has a strong influence on Agent $1$, in the end
Agent $2$ considers more plausible that $p$ is not ``too bad''. Finally, if we
consider $M_3$, where the agents have the same level of influence, the
system converges to 
a preference of approximately $0.35$ when $p\in \{1,2\}$ and $0.1$
otherwise. 

\begin{figure}[!t]
\centering
\begin{subfigure}{0.44\textwidth}
    \includegraphics[scale=0.3]{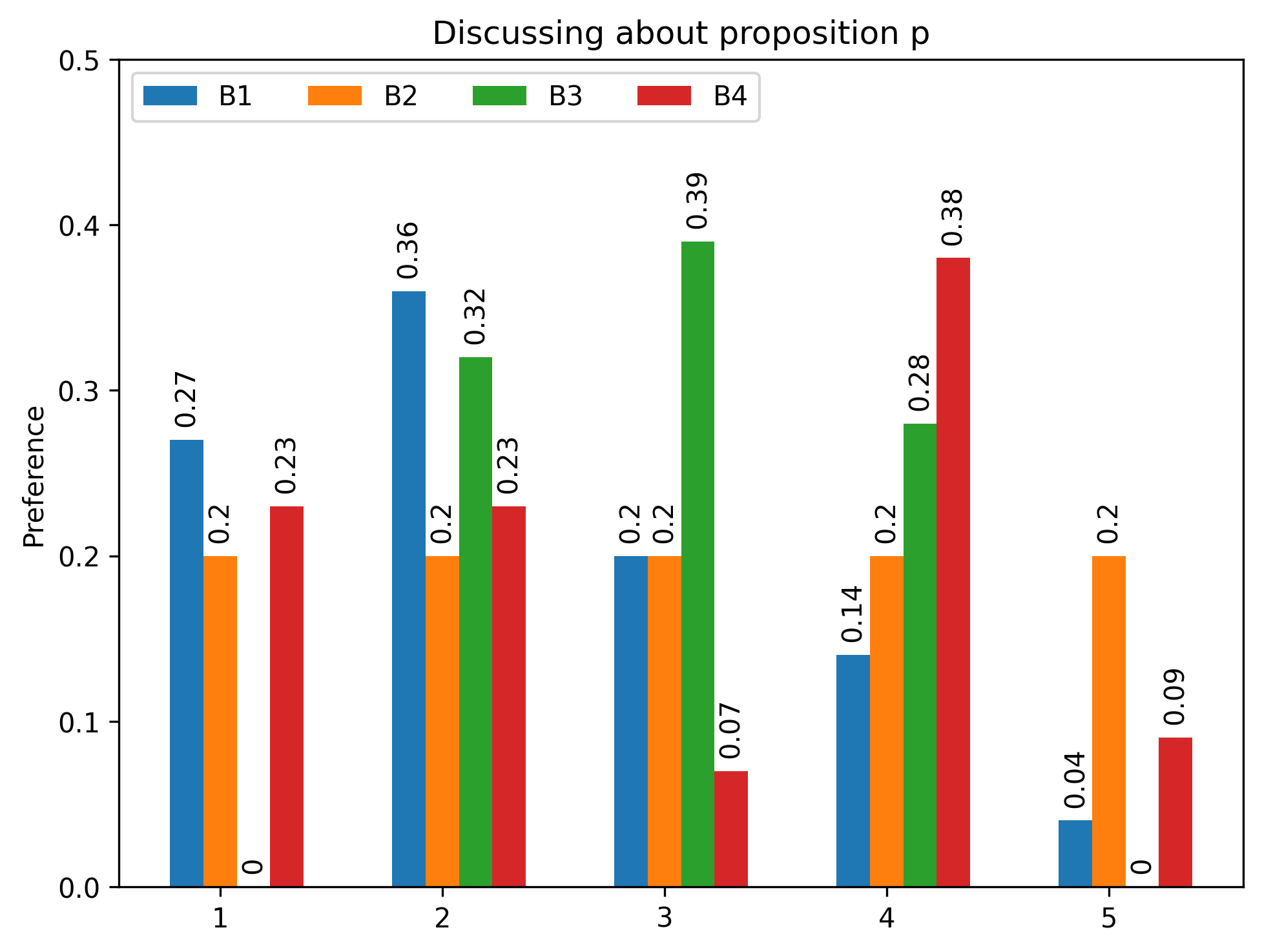}
    \caption{Resulting consensus when starting with the initial opinions in \Cref{ex:propp}.}
    \label{fig:con1}
\end{subfigure}
\begin{subfigure}{0.44\textwidth}
    \includegraphics[scale=0.3]{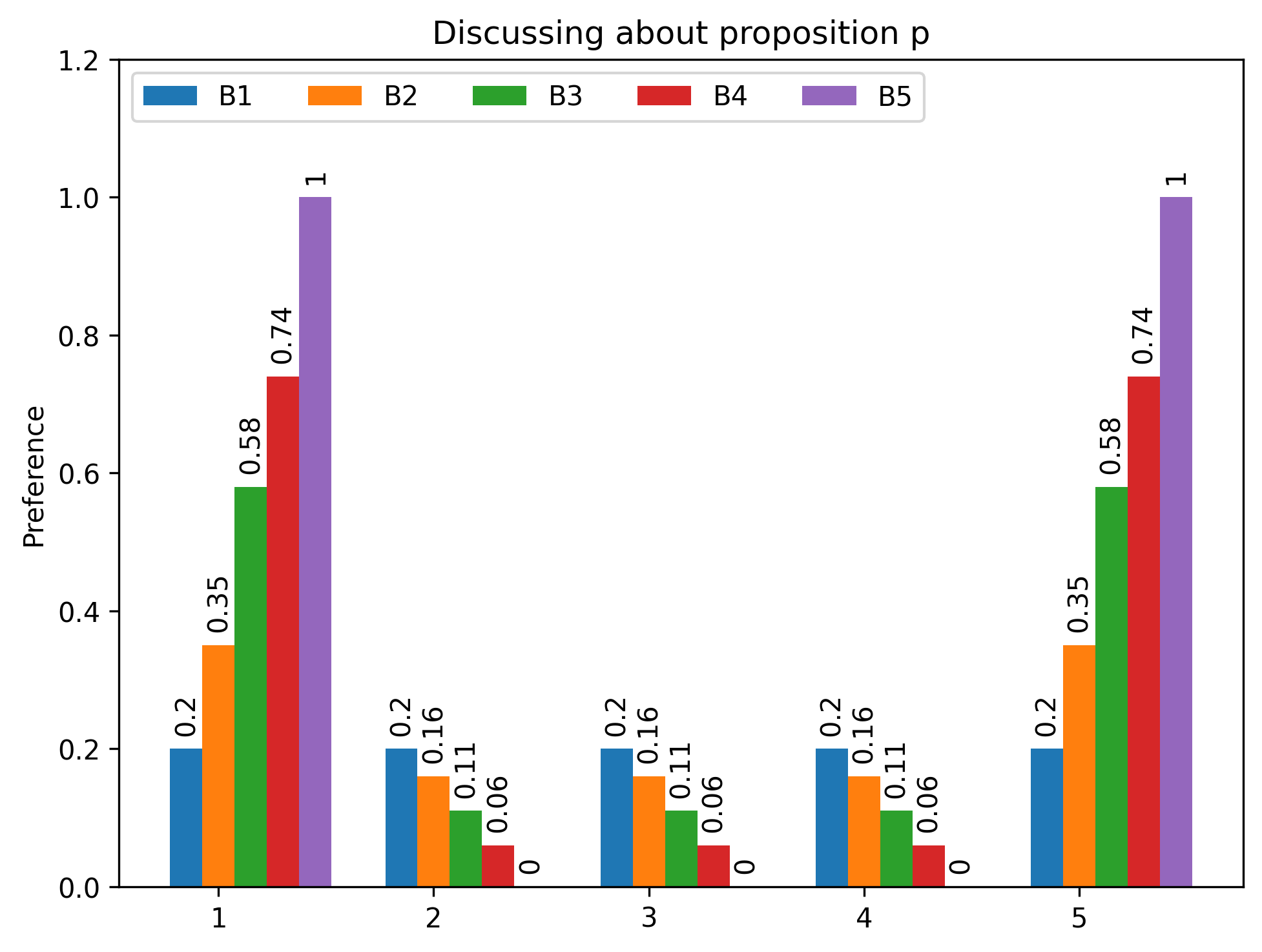}
    \caption{Resulting consensus when starting with the initial opinions in \Cref{ex:extremes}.}
    \label{fig:con2}
\end{subfigure}

    \caption{Preferences in the consensus when agents discuss a proposition $p$.}
\end{figure}

\begin{example}\label{ex:propp}

Let $\CInt{a,b}$ be as above and consider the following influence graph and
vectors of initial opinions, where agents are not completely sure how to assess
$p$ (\emph{partial information})

\[
\begin{array}{c}
M = \left[
	\begin{array}{llll}
     \CCte{0.2}& \CCte{0.3}& \CCte{0.4}& \CCte{0.1}\\
     \CCte{0.3}& \CCte{0.1}& \CCte{0.2}& \CCte{0.4}\\
     \CCte{0.5}& \CCte{0.1}& \CCte{0.2}& \CCte{0.2}\\
     \CCte{0.3}& \CCte{0.3}& \CCte{0.3}& \CCte{0.1}\\
	\end{array}
	\right]
\qquad
\begin{array}{lll}
    B_1 &=& [ \CInt{1,2}~~ \CInt{1,5}~~ \CInt{2,4}~~\CInt{1,3} ]\\
    B_2 &=& [ \CInt{1,5}~~ \CInt{1,5}~~ \CInt{1,5}~~\CInt{1,5} ]\\
    B_3 &=& [ \CInt{2,2}~~ \CInt{3,3}~~ \CInt{4,4}~~\CInt{3,3} ]\\
    B_4 &=& [ \CInt{1,2}~~ \CInt{1,3}~~ \CInt{4,4}~~\CInt{4,5} ]\\
\end{array}
\end{array}
\]

The consensus values obtained by starting with each of the initial opinions
$B_i$ above are shown in \Cref{fig:con1}. In $B_1$, the agents generally
assign a negative score to $p$, with a stronger preference for $p = 2$ (a value
present in all the initial beliefs). In $B_2$, the agents equally prefer all
possible values of $p$, achieving a consensus where no specific value is
favored over others. In $B_3$, the agents are more certain about the value of
$p$, with two of them believing that $p = 3$. Consequently, $p = 3$ becomes the
most preferred value in the consensus, followed by $p = 2$. In this scenario,
note also that none of agents believe that $p=1$ or $p=5$ and those values
receive a preference of $0$ in the consensus. In $B_4$, two agents are inclined
to assign a negative score to $p$, while the other two are more positive. Given
this particular influence graph, the system converges to a situation where $p =
4$ receives the highest preference.

\end{example}

The examples above consider opinions to be elements in $C_p$. Now 
we consider scenarios  where agent's opinions do not adhere to that
restriction. 

\paragraph{Preference for Extreme Positions.} The following example considers agents that prefer any extreme option than a moderate one. 
This kind of preference appears for example in risky decision-making, where an agent might prefer  high-risk, high-reward options over  safe middle-ground ones.
\begin{example}   Consider the following constraint

\[
	\CExt{x,y} \defsymbol \lambda p.  \left\{
			\begin{array}{lll}
				x & \mbox{ if } p=1 \mbox{ or } p=5\\
				y & \mbox{ otherwise } 
			\end{array}
			\right.
\]

Intuitively, the opinion $c_e = \CExt{1.0, 0.0}$ represents the situation where the agent ``likes
the extremes'' (regardless of which one), while the constraint $c_c=\CExt{0.0,
1.0}$ represents the opinion of an agent that prefers 
more moderate positions.

Consider the following situation (note that $M$ satisfies the conditions in \Cref{prop:conc})
\[
	\begin{array}{lllll}
	M = & \left[
			\begin{array}{lll}
				\CCte{0.3} && \CCte{0.7}\\
				\CCte{0.4} && \CCte{0.6}\\
		
			\end{array}
		  \right]
	& \quad & 
\begin{array}{lll}
B_1 &=&  [~c_e\quad d~] \\
B_2 &=&  [~c_e\quad c_c~] 
\end{array}
	\end{array}
\]
where $d[p\mapsto 2] =0.2$, $d[p\mapsto 3]=0.3$, and $d[p \mapsto v]=0.0$ otherwise. 
Starting from the initial beliefs $B_1$ and $B_2$ above, the system 
converges to the following opinions $c_1$ and $c_2$ respectively
\[
\begin{array}{lll}
c_1 &=& \{
1 \mapsto 0.36,
2 \mapsto 0.13,
3 \mapsto 0.19,
4 \mapsto 0.00,
5 \mapsto 0.36
\}
\\
c_2 &=& 
 \{
1 \mapsto 0.36,
2 \mapsto 0.64,
3 \mapsto 0.64,
4 \mapsto 0.64,
5 \mapsto 0.36
\}
\end{array}
\]

In $c_1$, the agents assign higher preferences (but less than the original 1.0
for Agent $1$) to the extreme positions. 
Since Agent $2$  has a stronger influence on Agent $1$, the opinion $c_2$ 
assigns higher preferences to the moderate positions than the extreme
ones. 
\end{example}

\begin{example}\label{ex:extremes}
Consider the influence graph $M$ in \Cref{ex:propp} and the  
initial opinions
\[
\begin{array}{lll l lll}
B_1 &=& [\CInt{1,5}~~ \CInt{1,5}~~ \CInt{1,5}~~ \CInt{1,5} \ ] & \qquad  & 
B_2 &=& [\CInt{1,5}~~ \CInt{1,5}~~ \CInt{1,5}~~ c_e \ ] \\
B_3 &=& [\CInt{1,5}~~ \CInt{1,5}~~ c_e, c_e \ ] & \qquad  & 
B_4 &=& [\CInt{1,5}~~ c_e~~ c_e~~ c_e \ ] \\
B_5 &=& [c_e~~ c_e~~ c_e~~ c_e \ ] 
\end{array}
\]

Starting with $B_1$, where none of the agents have a formed opinion about $p$ (constraint $\CInt{1,5}$),
we progressively add agents with ``extreme'' point of views (constraint $c_e$). For all these configurations, 
\Cref{fig:con2} shows the constraints obtained in the consensus. As expected, $B_1$ (respectively, $B_5$) is already
a consensus where no value of $p$ is more preferred than any other (respectively, the extreme values are equally preferred and the moderate ones are not considered). For this particular configuration of $M$, 
with $2$ ``extreme'' agents (opinion $B_3$), there is a clear tendency to prefer the valuations $p=1$ and $p=5$.
\end{example}

\paragraph{Discussing several topics.}
A constraint opinion model with one variable taking values from a finite domain
$D$ can be also interpreted as agents discussing different topics
simultaneously as the following example shows. 

\begin{example}\label{ex:topics}
Let  $D=\{1,2,3\}$ and consider the following
definition

\[
c_m(x_1,x_2,x_3) \defsymbol \lambda p. \left\{
\begin{array}{l}
 x_1 \mbox{ if p = 1}\\
 x_2 \mbox{ if p = 2}\\
 x_3 \mbox{ otherwise}
\end{array}
\right.
\]

The constraint $d = c_m(v_1,v_2,v_3)$ represents the opinion of an agent about
three different propositions, where the opinion about proposition $p_i$ is $d[p \mapsto i]$. As
an example, consider the situation

\[
\begin{array}{lll}
	M =  \left[
			\begin{array}{lll}
				\CCte{0.3} && \CCte{0.7}\\
				\CCte{0.5} && \CCte{0.5}
			\end{array}
			\right]	
\qquad
B_0 = [~c_m(0.3, 0.7, 0.1)\quad c_m(0.4, 0.4, 0.4)~]
\end{array}
\]

Agent $1$ is positive about the second proposition and tends to be more negative
about the other two propositions. In this case, the system converges to the following situation 
\[
c = \{
1 \mapsto 0.36,
2 \mapsto 0.52,
3 \mapsto 0.26
\}
\]

When non-constant constraints are considered in the influence graph, it is possible to
represent the situation where the influence of the agents depends on the
proposition being discussed. For example, consider the initial opinion $B_0$
above and the following influence graph 

\[ M =  \left[ 
 \begin{array}{lll}
c_m(0.3,0.2,0.1) && c_m(0.7,0.8,0.9)\\ 
c_m(0.5,0.1,0.8) && c_m(0.5,0.9,0.2) 
\end{array}
\right]	\]

Note that Agent $1$ exerts less influence over Agent $2$ 
when discussing about the second proposition  than when 
discussing the third proposition ($c_m(0.5,0.1,0.8)$). 
For this configuration, the agents converge to 
$
c = \{
1 \mapsto 0.36,
2 \mapsto 0.43,
3 \mapsto 0.26
\}
$. 
\end{example}

In the example above, the influence graph is not a matrix of
constant constraints; therefore, the consensus theorem on reals does not directly apply.
However, when the influences of the different topics are
independent as in the example above, we can naturally extend \Cref{prop:conc} as follows.

\begin{proposition}

Consider a constraint opinion model on $\bbR^+$  with one variable $p$ with finite domain
$D$ and an influence graph $M$. If for all $d\in D$ the matrix of constant constraints $M\{p
\mapsto d\}$ (the result of applying $\eta=\{p\mapsto d\}$ to each element
 in $M$) 
is row stochastic, strongly connected and aperiodic, then $\lim_{t\to\infty}M^t$
exists and all its rows 
are equal.
\end{proposition}

\subsection{Partial Information and Conditional Opinions}\label{app:cond}

\Cref{ex1} showed that using classical/crisp constraints (semiring $\bbB$), it is
possible to represent the situation where agents share partial information
about the actual value of a given variable. Moreover, using constraints as
influences, it is possible to ``filter'' the information coming from other
agents. In this section  we show how this idea can be further generalised when
$\bbB$ is replaced with $\bbR^+$.

Consider the situation where four agents are trying to decide the number of
members a selection committee must have. We represent that decision with a
variable $x$. Agents have different opinions about the actual value of $x$, and
such opinions can be naturally expressed as constraints. For instance,
the initial belief could be given by
\[
\begin{array}{llllll}
o_1: 4 \leq x\leq  6  &\qquad
o_2: 5 \leq x \leq 10 &\qquad
o_3: 6 \leq x \leq 7 &\qquad
o_4: 1 \leq x \leq 4 &\qquad
\end{array}
\]

The opinions above express different degrees of uncertainty, which are reflected on the
number of possible values $o_i$ allows. The initial belief of each agent is
a constraint $c_i$ such that $c_i[x\mapsto v] = 1$ if $o_i[v/x]$ is true and $c_i[x\mapsto v]=0$
otherwise. For instance, the constraint $\{4\leq x\leq 6\}$ is the function
that maps to $1$ the valuations that map $x$ to a value $v\in \{4,5,6\}$. 

The four agents need to take a decision and they define
how they will be influenced by the others' opinions, for instance

\[
M =
\left[ 
\begin{array}{ccccccc}
 	 {0.3} &  & {0.2} &  & {0.3} &  & {0.2}\\
     {0.2} &  & {0.3} &  & {0.1} &  & {0.4}\\
     {0.2} &  & {0.4} &  & {0.2} &  & {0.2}\\
     {0.1} &  & {0.1} &  & {0.5} &  & {0.3}
\end{array}
\right]
\]

 After some iterations, the system converges to the following constraint
\[
    \begin{array}{lll ll lll}
        c[x\mapsto v]&=&0.28 \mbox{ if $1\leq v \leq 3$} &\qquad&
        c[x\mapsto 4]&=&0.46\\
        c[x \mapsto 5]&=&0.44 & \qquad &
        c[x \mapsto 6]&=&0.72\\
        c[x \mapsto 7]&=&0.53 &\qquad&
        c[x \mapsto v]&=&0.25 \mbox{ if $8 \leq v \leq 10$} \\
        c[x \mapsto v]&=&0 \mbox{ otherwise}
    \end{array}
\]
 Hence, a selection committee of $6$ people
 is the best solution these agents can find that ``better'' satisfies their
 initial beliefs and influences.

Now suppose that the agents must also decide whether the selection committee
must include external members or not. Such a decision is modelled with a second variable $y$ with domain
$\{0,1\}$. Using constraints, it is natural to represent opinions such as
``with external members, I would prefer a committee of size $5\leq x \leq 6$
but, without them, I would prefer a committee of size $3 \leq x \leq 5$''. This
statement can be modeled as the implication $y=1\Rightarrow 5\leq x\leq 6$ (or
$y=0 \vee 5 \leq x \leq 6$) and $y=0 \Rightarrow 3 \leq x \leq 5$. We shall use
$c_s(5,6,3,5)$ to represent such an opinion where  $c_s$ is defined as
\[
    c_s(a_0,b_0, a_1,b_1) \defsymbol \lambda x~y.( (\{y=1\} + \{a_0 \leq x \leq b_0\}) \times (\{y=0\} +  \{a_1 \leq x \leq b_1\}))
\]

Consider also the definition
\[
    c_s'(s, a,b) \defsymbol \lambda x~y.( \{y=s\} \times   \{a \leq x \leq b\})
\]

The constraint $c_s'(1,4,6)$ represents the opinion ``the committee must
be formed by a number of $4\leq x\leq 6$ people and it must include external
members''. 

\begin{figure}[!t]
\centering
\begin{subfigure}{0.44\textwidth}
    \includegraphics[scale=0.3]{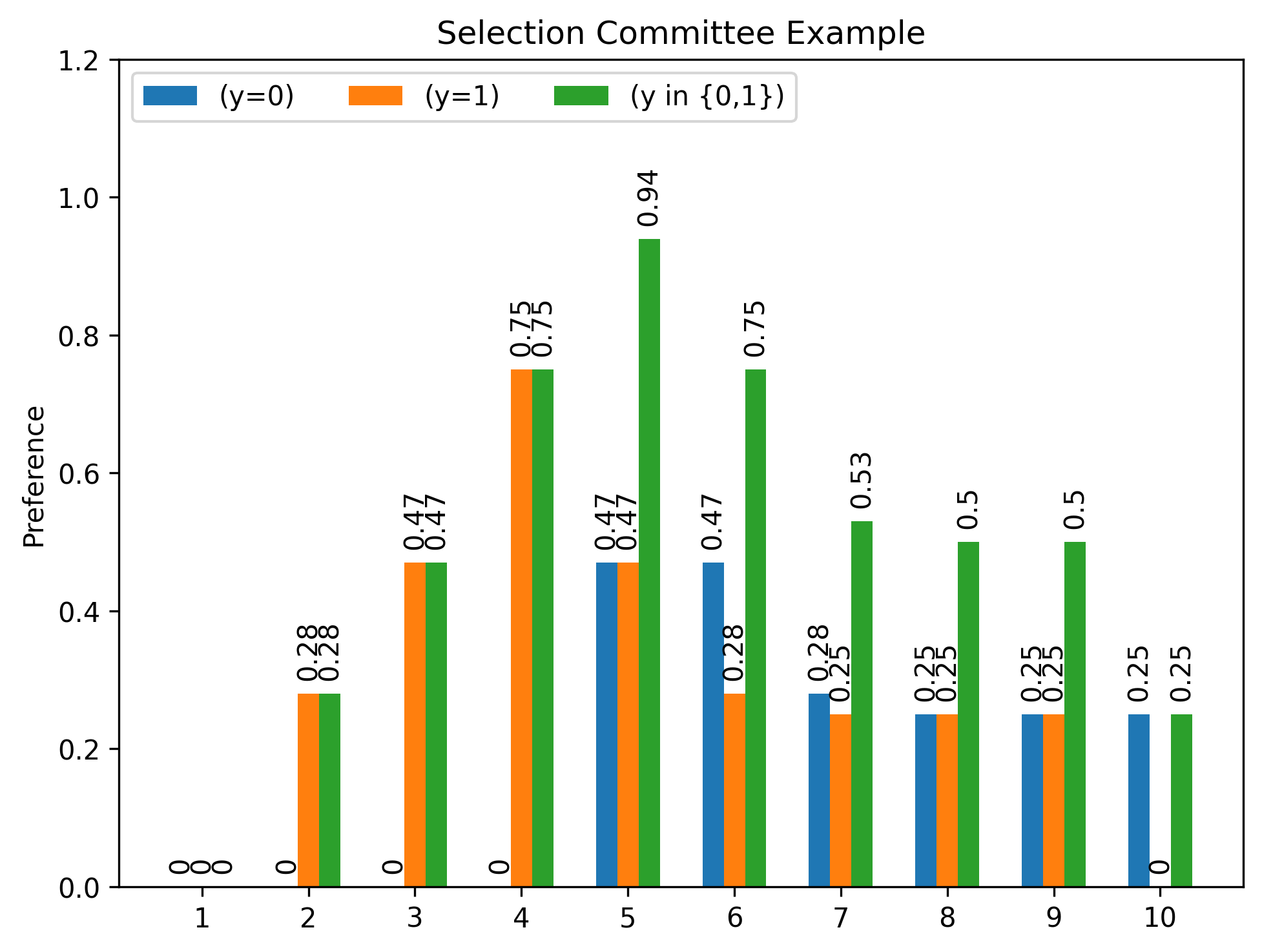}
    \caption{Preferences in the Consensus}
    \label{fig:sc1}
\end{subfigure}
\begin{subfigure}{0.44\textwidth}
    \includegraphics[scale=0.32]{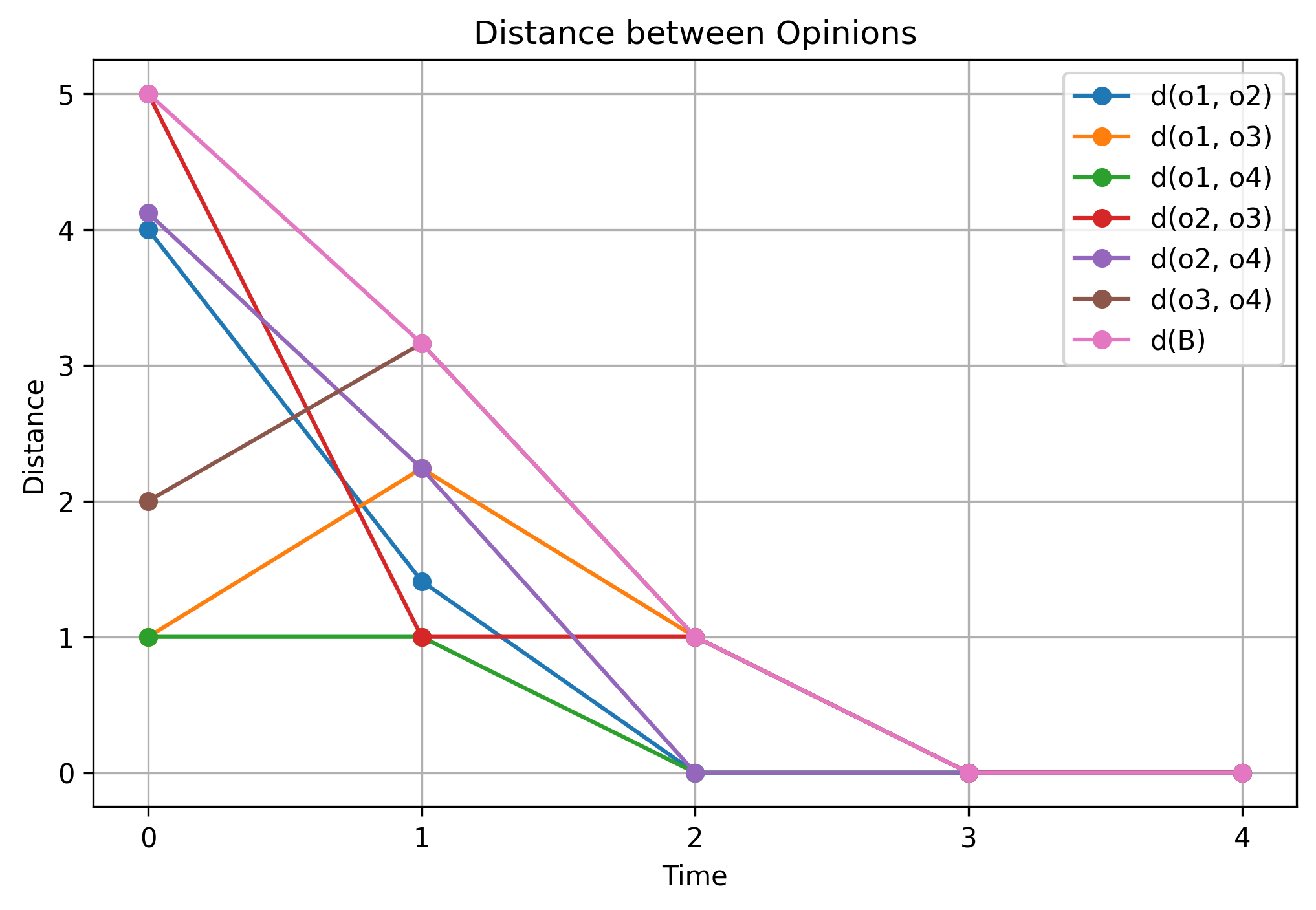}
    \caption{Distance between opinions (Ex. \ref{ex:dist})}
    \label{fig:sc2}
\end{subfigure}
\caption{Preferences in the consensus when agents discuss the size of a committee.}
\label{fig:sc}
\end{figure}

 \Cref{fig:sc1} shows the consensus $c$ reached with the influence graph $M$
above, when the initial vector of beliefs is
\[
    \begin{array}{lll l lll}
        B_0 &=& [~c_s(5,6,3,5)\quad c_s(8,10,7,9)\quad c_s(5,7,2,4)\quad c_s'(1,4,6)~] 
    \end{array}
\]

The green bars represent the opinion $\projC{c}{x}$ (the projection of $c$
onto the variable $x$). This projection captures the agents' preferences when
disregarding their opinions about the inclusion of external members. Similarly,
the constraint $d = \projC{c}{y}$ represents the preferences of the agents
regarding only the value of $y$. In this example, $d[y \mapsto 0] = 1.97$ and
$d[y \mapsto 1] = 3.0$. Thus, if we select the best assignment (constraint
$c$, blue and orange bars), the optimal choice is to form a selection committee
of 4 members, including external ones. However, if we ignore the decision about
external members ($\projC{c}{x}$, green bars), the preferred option is a
selection committee of 5 members. Finally, the constraint $\projC{c}{y}$
indicates that the agents prefer the inclusion of external members.

\subsection{Beliefs}\label{app:bel}
Hitherto the opinion of an agent $a\in A$ is a constraint of
type $c_a: (X \to D) \to S$. 
In this section we aim
to be more nuanced, and 
to be able to model 
the  beliefs of agents about themselves and about the other agents. 
To this end, a constraint should be of the form 
$c_a: A \to ((X \to D) \to S)$.
Consider e.g. the semiring $\bbB$ and the agents $A=\{a,b\}$. The
``constraint'' $c_a$ such that
$c_a(a)= \{ x > 42\}$
and
$c_a(b) = \{x\leq 20\}$ represents the situation
where $a$ thinks that $x > 42$, and $a$ thinks that $b$ believes that $x \leq 20$. 

Note that $A \to ((X \to D) \to S)$ is equivalent to
$(A \times (X \to D)) \to S$
and
$A \times (X \to D)$ is equivalent
to $(\{\bullet\} \to A) \times (X \to D)$, which in turn
is equivalent to the \emph{typed} 
functions
$(\{\bullet\} \uplus X) \to  (A \uplus D)$.
The results of the previous sections could be rephrased for constraints over typed valuations, even if we 
restrained from doing so for the sake of simplicity.
For the rest of this section 
we then consider valuations 
for typed variables $\{\bullet\} \uplus X$ and  
disjoint domains $A \uplus D$.
Also, we represent a 
constraint $c$ over such valuations
as $\oplus_{a \in A} c^a$ for 
$c^a\eta = c\eta$ if $\eta(\bullet) = a$ and $0$ otherwise.

For the influence graph, the elements  $M_{i,j}$ of the matrix are thus constraints that can be
also represented as  $M_{i,j} = \bigoplus_a M^a_{i,j}$, so that
$M = \bigoplus_a M^a$.
Since also a vector $B$ can be decomposed as
$\bigoplus_a B^a$, we have 
that $M B = \bigoplus_a M^aB^a$. 
We can further impose
the restriction that each component matrix is row stochastic, i.e. 
$M^a \1 = \1$. In this way, the results and definitions in
Section~\ref{update} can be recast to the case of beliefs.

Similar to \Cref{ex1}, consider the semiring $\bbB$, a set of two agents $A = \{a, b\}$ and the following constraints
and influence graphs
\[
\begin{array}{lll l lll}
c_a &=& \{x > 42, y \leq 30\}^a \oplus \{x \leq 20\}^b & \qquad &
c_b &=& \{x > 10, y = 20\}^b\\\\
	M &=& \begin{bmatrix}
		\TT^a \oplus \FF^b & \FF^a \oplus \TT^b  \\
		\TT^a \oplus \TT^b & \TT^a \oplus \TT^b
	\end{bmatrix} & \qquad &
	N &=& \begin{bmatrix}
		T & F  \\
		T & T
	\end{bmatrix} =
    \begin{bmatrix}
		T^a \oplus T^b & F^a \oplus F^b  \\
		T^a \oplus T^b & T^a \oplus T^b
	\end{bmatrix}
\end{array}
\]

Agent $a$ believes that $x>42$ and $y \leq 30$. Moreover, Agent $a$ thinks
that Agent $b$'s opinion is $x\leq 20$. On the other side, Agent $b$
thinks that $x>10$ and $y=20$, and Agent $b$ does not have any opinion about the values
Agent $a$ considers possible for $x$ and $y$. 

In $M$, note that  $M_{1,1}^b = \FF$ and $M_{1,2}^b = \TT$. This means that 
$a$ is willing to scrap its belief about $b$ and to accept the opinion of Agent $b$ about itself.
If we compose $M [~c_a\quad c_b~]$ we obtain
$[~d_a\quad d_b~]$ with
$$d_a = c_a^a \oplus c_b^b = \{x > 42, y \leq 30\}^a \oplus \{x > 10, y = 20\}^b$$
$$d_b = c_a \oplus c_ b = \{x > 42, y \leq 30\}^a \oplus \{x \leq 20\}^b \oplus
        \{x > 20, y = 20\}^b$$
where the latter equality is due to the fact that
$$\{x \leq 20\} \oplus
\{x > 10, y = 20\}
=
\{x \leq 20\} \oplus
\{x > 20, y = 20\}$$

Finally, 
composing $N [~c_a\quad c_b~]$ we obtain
$[~d_a\quad d_b~]$ with $d_a = c_a$ and $d_b = c_a \oplus c_b$.

\section{Measuring Opinion Difference}\label{sec:polarization}

Polarisation measures aim to quantify how divided a set of agents is concerning
their opinions \cite{ER:1994}. A key element in this assessment is the ground
distance between opinions, which serves as a way to determine whether
disagreements are minor or extreme. When opinions are exact values represented
as real numbers, as assumed in previous models for opinion dynamics, the
Euclidean distance and the absolute difference are natural choices.
Nevertheless, in the  model proposed here, opinions are not precisely known but
are instead represented by constraints. Hence,  defining a ground distance is more complex,
as we no longer compare single points but  uncertainty regions.

In this section we introduce a polarization measure for constraint opinion
models, aimed at quantifying the ``divergence'' between two constraints. The
key idea is to measure this divergence based on the distance between their
respective solution sets. Remember that, given a semiring value $s$,
$c^{-1}(s)$ is the set of assignments that an agent considers possible, or at least those
that are consistent with a ``preference level'' $s$. By evaluating this
distance, we gain insight into the degree of alignment or divergence between the
agents' opinions.

We fix a set of variables $X$, a domain of interpretation $D$, and a semiring
$\mathbb{S} = \langle S, \oplus, \0, \monop, \monid \rangle$. At first sight,
it seems natural to define the distance between constraints by
assuming that $\bbS$ is a metric space, equipped with a distance $\delta: S
\times S \to \bbR^+$. However, this allows to capture \emph{only}
the distance between two particular assignments 
but it does not tell us much about 
\emph{all} the possible assignments the agents consider plausible. Moreover, in
the case of the Boolean semiring, this measure will be the coarsest possible:
the distance is $0$ if 
$c_1\eta = c_2\eta$, or a value $\delta(\TT,\FF)=\delta(\FF,\TT)$ otherwise. 
The distance proposed here, instead, assumes that $D$ is a metric space 
that we lift to a metric space on valuations under the assumption
of a finite support for variables. 

If  $D$ is a metric space, equipped with a distance $\delta:D\times D \to \bbR^+$,   we can define a new metric space $D^n$ for any $n\in \bbN$, assuming the existence of a norm function 
$\lVert \cdot \rVert: \mathbb{R}^n \to \mathbb{R}^+$, so that we have
$$\delta((x_1, \ldots, x_n), (y_1, \ldots, y_n)) = \lVert \delta(x_1, y_1), \ldots, \delta(x_n, y_n) \rVert$$

The most used norm in $\mathbb{R}^n$ is the Euclidean norm $L^2$, so that we have
$$\lVert(z_1, \ldots, z_n)\rVert = \sqrt{z_1^2 +\ldots + z_n^2}$$

Consider for instance the examples in \Cref{sec:app},  where $D$ is a finite subset of $\mathbb{N}$.
We can consider the usual distance for $D$, namely 
$\delta(x, y) = \vert x - y \vert$. Adopting the $L^2$ norm for $\mathbb{R}^n$
results in the Euclidean distance on $\mathbb{N}^n$ for every $n$.

\begin{remark}
    In the rest of this section, we restrict constraints to be built over
    a  finite set $V$ of variables, and we assume that the domain $D$
    is a metric space, lifted to  
$D^{|V|}$.
\end{remark}

Given a semiring value $s$, the distance between two constraints $c$ and $d$ will be 
the  distance between the sets of assignments 
$c^{-1}(s)$ and $d^{-1}(s)$, thus comparing the ``solutions'' of the opinions
$c$ and $d$ with respect to a preference level 
$s$. Since $V\to D$ is a metric space, we can compare two sets of
assignments as follows. 

\begin{definition}[Hausdorff Distance]
    Let $\bbM$ be a metric space. The distance between an element of
    the metric space and a subset of elements of $\bbM$ is defined as 
    $\delta(v, B) = inf_{w \in B} \delta(v,w)$. 
    The forward distance between two subsets of $\bbM$ is defined as
     $
         \overrightarrow{\delta}(A, B) = sup_{v \in A} \delta(v,B)
     $. 
     For two non-empty sets $A$ and $B$,  the Hausdorff distance is defined as 
\[
    \delta_H(A,B) = \textit{max}\left(\overrightarrow{\delta}(A, B) ,\overrightarrow{\delta}(B, A) 
    \right)
\]
Moreover,  $\delta_H(A, \emptyset) = \delta_H(\emptyset,A)=\infty$ and
$\delta_H(\emptyset,\emptyset) = 0$. 
\end{definition}

 Intuitively, the distance $\delta_H(A,B)$  is the 
 greatest of all the distances from a point in the set $A$ to the closest point
 in the set $B$. Notice that this guarantees that any element of one set is
 within a distance of \emph{at most} $\delta_H(A,B)$ of some element of the
 other set. 

\begin{definition}[Distance between opinions]\label{def:dist}
 Let $s\in S$ and $c,d$ two constraints.
We define $\delta_s(c,d)=\delta_H(c^{-1}(s), d^{-1}(s))$ and
$\delta(c,d) = max\{ \delta_s(c, d) ~\mid~ s \in S  \}
$. 
Both definitions are well-given since $V$
is finite,
hence the images of $c$ and $d$ are so.
 \end{definition}

 \begin{remark}
    Note that the Hausdorff distance and its variants is often used to check the dissimilarity between two
    sets with respect to a third one, which is considered the set of the ground truths.
    In the case of $\delta_s$ we have a natural candidate, which is the constant constraint 
returning always $s$,
    whose  counterimage is 
    precisely $D^{\mid V \mid}$.
    Also note that
    for any $B \subseteq D^{\mid V \mid}$
    we have
    $\overrightarrow{\delta}(B, D^{\mid V \mid}) = 0$, and that $\overrightarrow{\delta}(D^{\mid V \mid}, B)$ intuitively tells the distance of the most extreme positions from those of $B$.
    Thus,
    dividing $\overrightarrow{\delta}(c^{-1}(s), d^{-1}(s))$ by $ \overrightarrow{\delta}(D^{\mid V \mid}, d^{-1}(s))$ can offer a (very rough) approximation of the dissimilarity between the most extreme solutions accepted by $c$ with value $s$ with respect to the solutions accepted by $d$
    with value $s$: if it is equal to $1$, then 
    some of those most extreme solutions in $D^{\mid V \mid}$ hold also in $c$. Alternatively, a more quantitative measure of similarity can be found simply by 
    dividing for the largest distance between two points in $D^{\mid V \mid}$.
\end{remark}

 These definitions can be generalized 
 to a finite set $O$ of constraints,
 so that e.g. 
$\delta_s(O) = max\{ \delta_s(c_i, c_j) ~\mid~ c_i,c_j\in O  \}$.
Such a measure can be interpreted as an indicator of
 \emph{polarisation} in the opinions of a set of agents. A higher value of $\delta_s(O)$
 corresponds to a greater distance between the two most \emph{antagonistic}
 agents. Moreover, small values of $\delta_s(O)$ indicate that agents tend to consider as
 plausible the same set of solutions.

 If the semiring $\bbS$ is equipped with an order $\leq$, we can define 
 $c^{-1}_{\geq}(s)$ as the  set of valuations 
$\{\eta~\mid~s \leq c\eta \}$, i.e. the valuations
that assign a value $s'$ at least as ``good'' as $s$. 
Accordingly, we define $\delta_{\geq s}(c,d)=\delta_H(c^{-1}_{\geq}(s), d^{-1}_{\geq}(s)) $
and, for a set of opinions $O$,  $\delta_{\geq s}(O) = max\{\delta_{\geq s}(c_i,c_j)~\mid~c_i,c_j \in O\}$. 

\begin{example}
    Consider the constraints in \Cref{ex1}, choosing $V = \{x, y\}$ and let 
\[
    \begin{array}{lll}
        A  &=& (c_1 \wedge c_2)^{-1}(\TT) = \{x \leq 42\} \cap \{y \leq 25\} \\
        B &=& (c_3 \vee c_4)^{-1}(\TT) = \{x \geq 15\} \cup \{y \geq 66\}
    \end{array}
\]
where $A$ (respectively $B$) is the set of valuations that
make the constraint $c_1\wedge c_2$ (respectively $c_3 \vee c_4$) true. 
In this example, 
$D = \{ 0, \ldots, 100\}$
 and we use $D^2$ as metric space, adopting the Euclidean distance.

 The intersection $A\cap B$ is  $\{15 \leq x \leq 42\} \cap \{y \leq 25\}$.
 Concerning the remaining elements of $A$, i.e. those satisfying $x < 15$, 
 the minimal distance with respect to $B$ is given by $15 - x$, and the maximal is for those elements laying on the line $x = 0$. 
 Thus, $\overrightarrow{\delta}(A, B) = 15$.
Now, for those elements of $B$ satisfying $x \leq 42$, the distance is either $0$ or $y-25$. Hence, it is maximal for those elements laying on the line $y = 100$. For those satisfying $y \leq 25$, the distance is either $0$ or $x - 42$. Hence, it is maximal for those elements laying on the line $x = 100$. For the remaining ones, i.e. those satisfying $x > 42$ and $y > 25$, the distance is given by
$\sqrt{(x-42)^2 + (y-25)^2}$. Clearly, in this case, it is maximal for the point $(100, 100)$ so that the overall forward distance is 
$\overrightarrow{\delta}(B, A) = \sqrt{(100-42)^2 + (100-25)^2} = \sqrt{8989} \approx 94.81$.
Hence, we have that $\delta(A, B) \approx 94.81$.

Consider now the constant constraint
that always returns $\TT$.
Once again, the distance is maximal for 
those elements lying on the line $x = 0$
such that $y \leq 51$,
witnessing $\overrightarrow{\delta}(D,B) = 15$
and $\overrightarrow{\delta}(A,B) \div \overrightarrow{\delta}(D,B) = 1$.
Instead, we have 
$max\{\delta(a,b) \mid a, b
\in D^2\} \approx 141.42$,
thus an approximation of similarity is given by 
$15 \div 141.42 \approx 10.6 \%$,
which is a proportionally small distance between the most extreme solutions.
\end{example}

\begin{remark}
Note that the considerations above also apply to the taxicab and the  Chebyshev distance. More precisely, $\overrightarrow{\delta}(A, B) = 15$, while $\overrightarrow{\delta}(B, A)$ is still given by the distance between the point $(100, 100)$ and $(42, 25)$, which are 
$\mid 100-42 \mid + \mid 100-25 \mid = 133$ and $m + max\{\mid 100-42 - m \mid, \mid 100-25-m \mid\} = 75$ for $m = min \{ \mid 100-42 \mid, \mid 100-25\mid \}$, respectively.
\end{remark}

\begin{example}\label{ex:dist}
    Consider the opinions in the selection committee example in 
    \Cref{app:cond}
\[
    \begin{array}{lll l lll}
        B^0 &=& [~c_s(5,6,3,5)\quad c_s(8,10,7,9)\quad c_s(5,7,2,4)\quad c_s'(1,4,6)~] 
    \end{array}
\]

\Cref{fig:sc2} shows the distances $\delta_{\geq s}(c_i,c_j)$ between agents $i$ and $j$ over time, 
along with the distance $\delta_{\geq s}(B^i)$ for the set of all the opinions when
$s=0.5$. Initially, Agents 2 and 3 exhibit the greatest divergence, whereas the
opinions of Agents 1 and 3, as well as Agents 1 and 4, are more closely
aligned. After the first interaction, the opinions of Agents 1 and 3 become
more distant compared to the initial state, a trend also observed between
Agents 3 and 4. Despite this, the overall distance $\delta_{\geq s}(B^i)$
decreases. After three iterations, all divergences disappear.

\end{example}

\section{Concluding Remarks}\label{sec:concluding}

We introduced the Constraint Opinion Model, a  generalisation of the
standard DeGroot model where opinions and influences are represented as soft
constraints rather than single real values. Our framework allows for modelling
belief revision scenarios involving partial information, uncertainty, and
conditional influences. We illustrated the expressiveness of our approach through
several examples and proposed a distance measure to quantify the difference between opinions where only partial information may be known.

\paragraph{Related Work.} 

There is a great deal of work on generalisations and variants of the DeGroot
model for more realistic scenarios (e.g.,
\cite{AlvimAKQV23,DBLP:conf/forte/AlvimSKV24,demarzo2003persuasion,chatterjee1977towards,Generalize2,Aranda2024}).
The work in \cite{AlvimAKQV23} extends the DeGroot model to capture agents
prone to confirmation bias, while \cite{DBLP:conf/forte/AlvimSKV24} generalises
\cite{AlvimAKQV23} by allowing agents to have arbitrary and differing cognitive
biases. The study in \cite{demarzo2003persuasion} introduces a version of the
DeGroot model in which self-influence changes over time, whereas influence on
others remains constant. The works in \cite{chatterjee1977towards,Generalize2}
explore convergence and stability, respectively, in models where influences
change over time. The study in \cite{Aranda2024} examines an asynchronous
version of the DeGroot model. Nevertheless, to our knowledge, no
generalisations of the DeGroot model address partial information.

This paper draws inspiration from the generalisations
of constraint solving and programming~\cite{cp-handbook} to deal with soft constraints
representing preferences, probabilities, uncertainty, or fuzziness. We build on
the \emph{semiring}-based constraint framework~\cite{jacm,jlamp17}, where
(idempotent) semirings define the operations needed to \emph{combine} soft
constraints and determine when a constraint (or solution) is \emph{better} than
another. Other similar frameworks exist, such as the \emph{valued} constraint
framework~\cite{DBLP:conf/ijcai/SchiexFV95}, which has been shown to be equally
expressive~\cite{DBLP:journals/constraints/BistarelliMRSVF99} to the one based
on semirings. 

Soft constraints~\cite{DBLP:journals/tocl/BistarelliMR06} have been used to
model agents or processes that share partial information in the style of
concurrent constraint
programming~\cite{cp-book,DBLP:journals/constraints/OlarteRV13}, where
processes can \emph{tell} (add) constraints to a common store of partial
information, and synchronise by \emph{asking} whether a constraint is entailed
by the current store. Timed extensions of this framework were proposed later
in~\cite{DBLP:conf/coordination/BistarelliGMS08}, where agents can tell and ask
different constraints along different time instants.
In~\cite{DBLP:journals/tplp/PimentelON14}, these languages have been shown to
have a strong connection with proof systems. To the best of our knowledge, this
is the first time that (soft) constraints have been used in the context of
belief revision.

\paragraph{Future Work.} We have laid the foundations to study DeGroot-based
belief revision under the lens of soft constraints, where opinions and
influences may include preferences, partial information and uncertainty.
There are several directions to continue this work. 
The most obvious one is try to recover the standard theory of consensus in a generic semiring. This issue is not straightforward. Consider e.g. two elements $X$ and
$Y$ such that $X \oplus Y = 1$ and 
$X \otimes Y = 0$, as they may be found in the free semiring given by the finite powerset. 
Now consider the matrix below
\[
	M =  \left[
			\begin{array}{lll}
				X & & Y\\
				Y && X \\
		
			\end{array}
		  \right]
\]
The matrix is involutory, as
e.g. $M_3$ in 
\Cref{ex1},
hence it never stabilises.
However, differently from $M_3$,
if both $X$ and $Y$ are different  from $0$, then the influence graph is strongly connected, hence an immediate generalisation of the 
consensus theorems does not hold. We believe that we could obtain it for semirings without zero divisors, along the line of the results on selective-invertible dioids presented in~\cite{dioids}.
Second, it is worth exploring  other models for social learning where, differently 
from the DeGroot model, not all
the agents interact at the same time but only two (as in the  gossip 
model~\cite{Fagnani2008}) or some  arbitrary set of them (as in the hybrid
model in~\cite{DBLP:conf/wrla/OlarteRRV24}). This is specially interesting when
agents may discuss about different topics, as shown in \Cref{ex:topics}. We also
plan to extend the rewriting logic based framework  proposed
in~\cite{DBLP:conf/wrla/OlarteRRV24} to the model proposed here, thus
providing a framework for the (statistical) analysis of systems of agents discussing soft
constraints. Third, the  model proposed here  allows for partial information about opinions and influences. Extending the DeGroot model with \emph{partial information}  about
cognitive biases seem a natural line of future research.


\begin{thebibliography}{10}
\providecommand{\url}[1]{\texttt{#1}}
\providecommand{\urlprefix}{URL }
\providecommand{\doi}[1]{https://doi.org/#1}

\bibitem{DBLP:conf/forte/AlvimAKQV21}
Alvim, M.S., Amorim, B., Knight, S., Quintero, S., Valencia, F.: A multi-agent
  model for polarization under confirmation bias in social networks. In:
  Peters, K., Willemse, T.A.C. (eds.) FORTE 2021. Lecture Notes in Computer
  Science, vol. 12719, pp. 22--41. Springer (2021)

\bibitem{AlvimAKQV23}
Alvim, M.S., Amorim, B., Knight, S., Quintero, S., Valencia, F.: A formal model
  for polarization under confirmation bias in social networks. Logical Methods
  in Computer Science  \textbf{19}(1) (2023)

\bibitem{fc}
Alvim, M.S., Knight, S., Valencia, F.: Toward a formal model for group
  polarization in social networks. In: Alvim, M.S., Chatzikokolakis, K.,
  Olarte, C., Valencia, F. (eds.) The Art of Modelling Computational Systems.
  Lecture Notes in Computer Science, vol. 11760, pp. 419--441. Springer (2019)

\bibitem{DBLP:conf/forte/AlvimSKV24}
Alvim, M.S., da~Silva, A.G., Knight, S., Valencia, F.: A multi-agent model for
  opinion evolution in social networks under cognitive biases. In: Castiglioni,
  V., Francalanza, A. (eds.) FORTE 2024. Lecture Notes in Computer Science,
  vol. 14678, pp. 3--19. Springer (2024)

\bibitem{Aranda2024}
Aranda, J., Betancourt, S., D{\'{\i}}az, J.F., Valencia, F.: Fairness and
  consensus in an asynchronous opinion model for social networks. In: Majumdar,
  R., Silva, A. (eds.) CONCUR 2024. Leibniz International Proceedings in
  Informatics, vol.~311, pp. 7:1--7:17. Schloss Dagstuhl - Leibniz-Zentrum
  f{\"{u}}r Informatik (2024)

\bibitem{DBLP:conf/coordination/BistarelliGMS08}
Bistarelli, S., Gabbrielli, M., Meo, M.C., Santini, F.: Timed soft concurrent
  constraint programs. In: Lea, D., Zavattaro, G. (eds.) COORDINATION. Lecture
  Notes in Computer Science, vol.~5052, pp. 50--66. Springer (2008)

\bibitem{jacm}
Bistarelli, S., Montanari, U., Rossi, F.: Semiring-based constraint
  satisfaction and optimization. Journal of the ACM  \textbf{44}(2),  201--236
  (1997)

\bibitem{DBLP:journals/tocl/BistarelliMR06}
Bistarelli, S., Montanari, U., Rossi, F.: Soft concurrent constraint
  programming. ACM Transactions on Computational Logic  \textbf{7}(3),
  563--589 (2006)

\bibitem{DBLP:journals/constraints/BistarelliMRSVF99}
Bistarelli, S., Montanari, U., Rossi, F., Schiex, T., Verfaillie, G., Fargier,
  H.: Semiring-based csps and valued csps: Frameworks, properties, and
  comparison. Constraints  \textbf{4}(3),  199--240 (1999)

\bibitem{chatterjee1977towards}
Chatterjee, S., Seneta, E.: Towards consensus: Some convergence theorems on
  repeated averaging. Journal of Applied Probability  \textbf{14}(1),  89--97
  (1977)

\bibitem{Generalize2}
Chen, Z., Qin, J., Li, B., Qi, H., Buchhorn, P., Shi, G.: Dynamics of opinions
  with social biases. Automatica  \textbf{106},  374--383 (2019)

\bibitem{Degroot1974}
DeGroot, M.H.: Reaching a consensus. Journal of the American Statistical
  association  \textbf{69}(345),  118--121 (1974)

\bibitem{demarzo2003persuasion}
DeMarzo, P.M., et~al.: Persuasion bias, social influence, and unidimensional
  opinions. The Quarterly Journal of Economics  \textbf{118}(3),  909--968
  (2003)

\bibitem{ER:1994}
Esteban, J., Ray, D.: {On the Measurement of Polarization}. Econometrica
  \textbf{62}(4),  819--851 (1994)

\bibitem{Fagnani2008}
Fagnani, F., Zampieri, S.: Asymmetric randomized gossip algorithms for
  consensus. IFAC Proceedings Volumes  \textbf{41}(2),  9051--9056 (2008)

\bibitem{jlamp17}
Gadducci, F., Santini, F., Pino, L.F., Valencia, F.D.: Observational and
  behavioural equivalences for soft concurrent constraint programming. Journal
  of Logic and Algebraic Methods in Programming  \textbf{92},  45--63 (2017)

\bibitem{Golup2017}
Golub, B., Sadler, E.: Learning in social networks. SSRN  (2017)

\bibitem{dioids}
Gondran, M., Minoux, M.: Graphs, Dioids and Semirings. Springer (2008)

\bibitem{Hansson-2001}
Hansson, S.O., Fermé, E.L., Cantwell, J., Falappa, M.A.: Credibility limited
  revision. Journal of Symbolic Logic  \textbf{66}(4),  1581--1596 (2001)

\bibitem{Makinson1997-MAKSR}
Makinson, D.: Screened revision. Theoria  \textbf{63}(1-2),  14--23 (1997)

\bibitem{DBLP:conf/wrla/OlarteRRV24}
Olarte, C., Ram{\'{\i}}rez, C., Rocha, C., Valencia, F.: Unified opinion
  dynamic modeling as concurrent set relations in rewriting logic. In: Ogata,
  K., Mart{\'{\i}}{-}Oliet, N. (eds.) WRLA 2024. Lecture Notes in Computer
  Science, vol. 14953, pp. 104--123. Springer (2024)

\bibitem{DBLP:journals/constraints/OlarteRV13}
Olarte, C., Rueda, C., Valencia, F.D.: Models and emerging trends of concurrent
  constraint programming. Constraints  \textbf{18}(4),  535--578 (2013)

\bibitem{paz:hal-04918975}
Paz, J., Rocha, C., Tob{\`o}n, L., Valencia, F.: Consensus in models for
  opinion dynamics with generalized-bias. In: Cherifi, H., Donduran, M., Rocha,
  L., Cherifi, C., Varol, O. (eds.) COMPLEX NETWORKS 2024. Studies in
  Computational Intelligence, vol.~1188, pp. 340--351. Springer (2025)

\bibitem{DBLP:journals/tplp/PimentelON14}
Pimentel, E., Olarte, C., Nigam, V.: A proof theoretic study of soft concurrent
  constraint programming. Theory and Practice of Logic Programming
  \textbf{14}(4-5),  649--663 (2014)

\bibitem{cp-handbook}
Rossi, F., van Beek, P., Walsh, T. (eds.): Handbook of Constraint Programming,
  vol.~2. Elsevier (2006)

\bibitem{cp-book}
Saraswat, V.A.: Concurrent Constraint Programming. MIT Press (1993)

\bibitem{DBLP:conf/ijcai/SchiexFV95}
Schiex, T., Fargier, H., Verfaillie, G.: Valued constraint satisfaction
  problems: Hard and easy problems. In: IJCAI (1). pp. 631--639 (1995)

\end{thebibliography}
\end{document}